\documentclass[12pt]{article}
\usepackage{amssymb}
\usepackage{epsfig}
\usepackage{amsmath}
\begin{document}


\begin{flushright}
hep-th/0409092\\
CITA-2004-16
\end{flushright}

\bigskip\bigskip

\begin{center}
{\Large \bf  Folding Branes}
\end{center}

\bigskip\bigskip

\renewcommand{\thefootnote}{\fnsymbol{footnote}}

\centerline{\bf Omid
Saremi$^1$\footnote{omidsar@physics.utoronto.ca}} \centerline{\bf
Lev Kofman$^{2,3}$\footnote{kofman@cita.utoronto.ca}}
\centerline{\bf Amanda W.
Peet$^{1,3}$\footnote{peet@physics.utoronto.ca}}
\smallskip
\medskip
\centerline{\it $^1$  Department of Physics} \centerline{\it $^2$
Canadian Institute for Theoretical Astrophysics} \centerline{\it
University of Toronto} \centerline{\it 60 St.\! George Street}
\centerline{\it Toronto, Ontario} \centerline{\it Canada M5S 1A7}
\medskip
\centerline{\it $^3$ Cosmology and Gravity Program} \centerline{\it
Canadian Institute for Advanced Research}
\bigskip

\setcounter{footnote}{0}
\renewcommand{\thefootnote}{\arabic{footnote}}

\bigskip\bigskip
\bigskip\bigskip

\abstract{We study classical dynamics of a probe D$p$-brane moving in
a background sourced by a stack of D$p$-branes. In this context the
physics is similar to that of the effective action for open-string
tachyon condensation, but with a power-law runaway potential. We show
that small inhomogeneous ripples of the probe brane embedding grow
with time, leading to folding of the brane as it moves. We give a
full nonlinear analytical treatment of inhomogeneous brane dynamics,
suitable for the Dirac-Born-Infeld + Wess-Zumino theory with
arbitrary runaway potential, in the case where the source branes are
BPS. In the near-horizon geometry, the inhomogeneous brane motion has
a dual description in terms of free streaming of  massive
relativistic test particles originating from the initial hypersurface
of the probe brane. We discuss limitations of the effective action
description around loci of self-crossing of the probe brane
(caustics).  We also discuss the effect of brane folding in
application to the theory of cosmological fluctuations in string
theory inflation.}

\vfill

\noindent September 2004.

\setcounter{page}{0}
\newpage
\section{Introduction}
In this paper we investigate generic inhomogeneous solutions
describing a probe D$p$-brane or anti-D$p$-brane falling in the
near-horizon D-brane background geometry. A particularly interesting
case occurs for D3-branes, where the background involves anti de
Sitter space. The background geometry is produced by the stack of $N$
source D$p$-branes. We will show that dynamics  of brane motion can
be reduced to that of the Dirac-Born-Infeld (DBI) type, with the
action
\begin{equation}\label{action}
S = -\int d^{p+1} x \, V(T) \sqrt{1 + \partial_\mu T
\partial^\mu T} \ .
\end{equation}
Here $T(x^{\mu})$ is a scalar field (dimensionless: in units of
$\alpha'$) and $V(T)$ is its runaway potential with no minima. The
action should be understood in the truncated approximation. While
first derivatives can be arbitrarily large, the action is valid only
in the regime where second and higher derivatives are small. The
action (\ref{action}) is identical to the effective field theory
action of the open string theory tachyon describing unstable non-BPS
D-branes as found in e.g.\footnote{We do not attempt to be exhaustive
referencing the string field theory literature here.}
\cite{Garousi,Sen1,Bergshoeff}. The similarity between the DBI
effective action for the tachyon and that of a probe D-brane near
NS5-Branes, for homogeneous solutions, was recently studied by
Kutasov \cite{kutasov1}.  Some other recent works on similar topics
(again for the homogeneous case) include
\cite{Panigrahi,Yavartanoo,Ghodsi,Kutasov2,Sahakyan}.

Previous works \cite{fks,felderkofman} on the generic inhomogeneous
solution in the theory (\ref{action}) were focused on the open string
theory tachyon effective action, with runaway potential of
exponential asymptotic form: $V(T) \simeq \exp(-T)$. A generic
analytic solution was found \cite{felderkofman} which, in particular,
shows the growth of $T$ inhomogeneities resulting in tachyon energy
fragmentation and folding of $T$. In our paper here, we extend the
method of \cite{felderkofman} to treat generic inhomogeneities for a
test D-brane or anti-D-brane freely falling in a background D-brane
geometry.  Folding of the field $T$ in this context acquires a
transparent interpretation, as the folding of the brane embedded in
the higher dimensional spacetime background. Folding of an
inhomogeneous probe D-brane should also transpire for the background
geometry generated by source NS5-Branes.

The setup of moving a brane in a background spacetime generated by
many branes is often an ingredient of string theory cosmological
construction including models of inflation, as in e.g.
\cite{evaat,Silverstein2}. Usually generation of inflationary
perturbations in these models are treated in the framework of
four-dimensional effective theory. The effect of the brane folding
which we find here is based on the full treatment of the probe brane
dynamics. Thus it should be incorporated in the complete treatment of
fluctuations in cosmological applications.

Before proceeding with the story of an inhomogeneous probe D-brane in
the background geometry, let us recall the story of the inhomogeneous
rolling tachyon. Study of the rolling homogeneous tachyon in
classical string field theory has shown that, at late times, the
system is described by a pressureless fluid ``tachyon matter'', with
no perturbative open string modes at the bottom of the potential.
Turning on $g_{s}$ would lead to subsequent decay of the unstable
brane, mainly to massive closed string modes as in \cite{Lambert}.
According to the proposal of \cite{Sen2}, this process has a dual
description purely in terms of the rolling open string tachyon.

It is difficult to find exact solvable CFTs describing open string
tachyon dynamics for a {\it generic} inhomogeneous tachyon profile.
Some progress in this regard was made concerning evolution of the
string theory tachyon with a plane wave tachyon profile
\cite{larsennt}. In this case, the tachyon decays into equidistant
plane-parallel singular hypersurfaces of codimension one, which were
interpreted as kinks. This inhomogeneous profile is atypical, though,
in the sense that fragmentation between kinks does not occur. In the
general case, based on cosmological intuition, we expect both types
of structures: weblike fragmentation and topological defects.

Quite apart from progress in the CFT approach, the action
(\ref{action}) has in fact proven to capture many striking features
of tachyon condensation. The relatively simpler formulation of
tachyon dynamics, in terms of the effective action (\ref{action}),
has therefore triggered significant interest in investigation of the
field theory of the tachyon, the possible role of tachyons in
cosmology, and so forth. Indeed, the end point of string theory brane
inflation is annihilation of D-branes and anti-D-branes, which leads
to the formation and subsequent fragmentation of a tachyon condensate
\cite{fks,felderkofman}. So the potential role of the tachyon in
cosmology cannot be understood without first understanding its
fragmentation.

In \cite{fks}, it was found  that the evolution of the tachyon field
$T(t, \vec x)$ can be viewed as a mapping $T(t_0, \vec x_0) \to T(t,
\vec x)$, that evolves to become multi-valued, with singularities at
caustics generated. This seemed to be a generic behavior for runaway
potentials.  More recently, a study \cite{felderkofman} considered
the generic {\it inhomogeneous} tachyon field, in the region where it
rolls down one side of the potential. Formation of sharp features in
the tachyon energy density, due to fragmentation, was observed. The
tachyon energy density pattern is reminiscent of the illumination
pattern at the bottom of a swimming pool, or the web-like large scale
structure of the universe. The similarity is not coincidental: the
underlying mathematics has common features in all three cases. In the
free streaming approximation of \cite{fks} this is just focusing of
particle trajectories corresponding to higher density concentrations
and, further, to the formation of caustics at the loci where
trajectories cross. These features, which are related to the
convergence of characteristics of the field $T$, are distinguished
from topological defects. The full picture must incorporate both
effects: formation of kinks and tachyon fragmentation in the space
between them.  In the context of brane inflation ending with
annihilation of D-branes and anti-D-branes, the tachyon is a complex
field and strings will be created. The web-like structure of
fragmentation which will be superposed with the network of strings.

In this paper, we look at another situation, concerning the radial
motion of a probe brane in particular string theory backgrounds.
Interestingly, moving probe D-branes in the background of RR-charged
or solitonic branes of various dimensions gives rise to an effective
probe dynamics in which the DBI part, after some field redefinition,
has the same structure as (\ref{action}) -- with a runaway potential
of {\it power law} type, i.e.
\begin{equation}\label{poten}
V(T)=c \, T^{-\alpha}
\end{equation}
 for some $\alpha>0$ \footnote{Clearly, this behavior differs from that
of the usual open string tachyon potential, which is exponentially
suppressed for large values of tachyon field.}. For example, in
the recent work \cite{kutasov1}, dynamics of a BPS test D-brane in
the vicinity of stack of NS5-branes was considered, and for radial
motion of the probe an action of type (\ref{action}) arises, with
a `tachyon' potential falling off exponentially for large value of
the `tachyon' field. In general, the DBI part of the probe action
for BPS D$p$-branes probing a stack of other branes in the
near-horizon region of the background has dynamics of type
(\ref{action}), but with a runaway potential (\ref{poten}) with
index $\alpha\equiv 2(7-p)/(5-p)$, which is positive, and hence
sensible, for $0\leq p<5$.

Clearly, in these cases, there is a spacetime interpretation of the
`condensation' of the radial `tachyon' mode, in terms of motion of
the probe brane in a background geometry. Analysis of such may open
a window to a better understanding of the full open string tachyon
condensation itself \cite{kutasov1}.

In cosmological applications, the radial distance $r$  to the probe
brane plays the role of the inflaton field \cite{dvalitye}. In a
particular cosmological setup,  cosmological scenarios based on the
motion of a test anti-D3-brane in the throat region of a large number
of D3-branes have recently been studied \cite{evaat,Silverstein2}.
Dynamics of the radial mode \footnote{Apart from its coupling to the
4D gravity, that is.} is given by an action of the above type
(\ref{action}) with a power-law runaway potential with $\alpha=4$.

Inhomogeneous fluctuations of this field during inflation generate
primordial cosmological fluctuations. Therefore it is important to
know the time evolution for not only homogeneous, plane parallel
brane motion in the anti de Sitter or more general background
geometry, but generic inhomogeneous branes.

Understanding of the generic solution of the DBI type effective
action may give insight to the structure of the theory.

The plan of our paper in the following. In Section
\ref{sec:probebrane} we introduce the supergravity description of the
probe brane freely falling in the background geometry. We will show
that the brane dynamics is reduced to the compact form of the DBI
plus WZ effective action for a single scalar $T$ with the power law
potential. In Section \ref{sec:freestream} we make the first
approximation towards the inhomogeneous solution of the non-linear
equation of motion for $T$, the free-streaming approximation.  Yet,
the free-streaming approximation is not enough to calculate the
stress-energy tensor of the brane. In Section \ref{sec:fullsoln} we
construct the full analytic solution in terms of asymptotic series
expansion for $T$. Most importantly, the next-to-leading term allows
to calculate the stress-energy of the brane. The answer is
transparent and allows a conjecture of a duality between a probe
brane freely falling into the near-horizon background geometry (such
as the AdS throat of D3-branes) and a collection of massive
relativistic particles, see Section \ref{sec:discussion}. In this
final section we also discuss various aspects of the physics of the
folding branes and its cosmological applications.

\section{Probe brane in background brane
geometry}\label{sec:probebrane}
The physics of a probe D$p$-brane or anti-D$p$-brane moving in the
background of $N$ source D$p$-branes is given by the usual
string-frame DBI plus Wess-Zumino action,
\begin{equation}\label{dbiwzaction}
S_{\rm probe}=S_{\rm DBI} + S_{\rm WZ} = -\tau_{p}\int
d^{p+1}\sigma e^{-\Phi} \sqrt{-\det(\mathbb{P}(g_{\alpha\beta))}}
\mp \tau_p\int \mathbb{P}(C_{\mathit (p+1)})
\end{equation}
where $\mp$ denotes - for probe anti-branes or + for branes, and
$\mathbb{P}$ is for pullback.

\begin{figure}
\begin{center}
\epsfxsize=.75\columnwidth \epsfbox{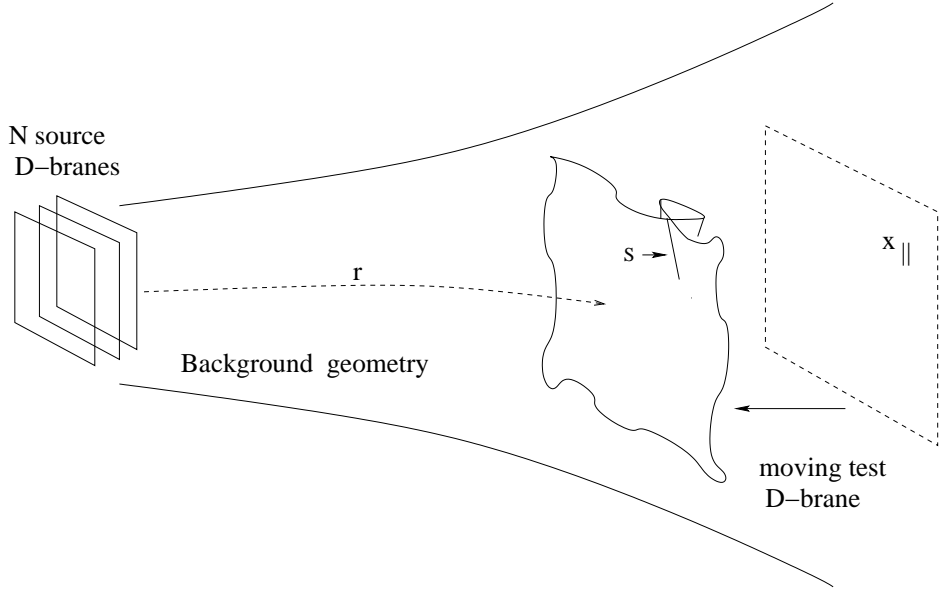} \caption{Sketch of
probe brane moving in the background geometry.} \label{fig:brane}
\end{center}
\end{figure}

For a (non-extremal) D$p$-brane spacetime in string frame, in a
standard coordinate system,
\begin{eqnarray}\label{dpsugra}
e^\Phi&=&H^{(3-p)/4}\nonumber\\
ds^2&=& H^{-1/2}\left(-Kdt^2+dx_{\parallel}^2\right) +
H^{1/2}\left(dr^2/K+r^2d\Omega_{8-p}^2\right) \nonumber\\
C_{\mathit (p+1)} &=& \zeta^{-1}\left(1-H^{-1}\right) dt\wedge
dx_1\cdots \wedge dx_{p}
\end{eqnarray}
The harmonic function is given by
\begin{equation}\label{hpzeta}
H=1 + \zeta {\frac{g_s N \ell_s^{7-p}}{r^{7-p}}} \equiv 1+
\left({\frac{R}{r}}\right)^{7-p} \,
\end{equation}
where $r^2=\sum_{i}(x_\perp^i)^2$. The event horizon is signalled by
$K(r)\rightarrow 0$, with
\begin{equation}\label{kofrx}
K(r)= 1- {\frac{r_H^{7-p}}{r^{7-p}}}\equiv 1-(2x)\left(H-1\right)
\end{equation}
where
\begin{equation}
x \equiv {\frac{1}{2}}\left({\frac{r_H}{R}}\right)^{7-p}
={\frac{(1-\zeta^2)}{2\zeta}}\quad {\rm or} \quad \zeta =
\sqrt{1+x^2}-x\,.
\end{equation}
$\zeta\in (0,1]$ encodes the degree of non-extremality.

It follows easily that
\begin{equation} S_{\rm probe} = -\tau \int
d^{p+1}\sigma {\frac{1}{H}} \left[
\sqrt{-\det(H{\bar{g}}_{ij}\partial_\alpha x_\perp^i\partial_\beta
x_\perp^j + {\bar{g}}_{\mu\nu}\partial_\alpha
x_\parallel^\mu\partial_\beta x_\parallel^\nu)} \pm \zeta
{\frac{\partial(x_\parallel)}{\partial(\sigma)}} \right]
\end{equation}
where for convenience we use an auxiliary metric
$({\bar{g}}_{\mu\nu})\equiv (-K,1,\ldots 1,1/K,r^2\ldots r^2)$, which
collapses to the flat metric in the BPS limit. In static gauge, where
$\sigma^{a}=x^{a}$ for $a=0,\ldots,p$, this expression simplifies in
the usual way to give
\begin{equation}\label{sprobestatic}
S_{\rm probe}^{\rm (static)}=-\tau \int d^{p+1}x_\parallel
{\frac{1}{H}} \left[\sqrt{1+H{\bar{g}}_{ij}\partial_\alpha
x_\perp^i\partial_\beta x_\perp^j{\bar{g}}^{\alpha\beta}} \pm \zeta
\right]
\end{equation}
A sketch of the probe brane in the background sourced by a stack of
$N$ D$p$-branes is shown in Figure~\ref{fig:brane}.

Now let us restrict to configurations in which the transverse motion
is purely radial. In this case, the physics simplifies to that of
just one field describing the motion of the test (probe) brane. This
can be seen by performing a field redefinition to give a `tachyon'
\begin{equation}\label{threlation}
T={\frac{R}{(p-7)}}\int dH \sqrt{\frac{H}{K(H)}}(H-1)^{-(8-p)/(7-p)}
\,.
\end{equation}
Physically, this expression (\ref{threlation}), in combination with
the expression (\ref{hpzeta}) for $H(r)$, gives the explicit
connection of the radial coordinate $r$ and the field $T$. In other
words, $T$ has clear geometrical interpretation in terms of the
radial position of the brane embedding as $r=r(t, \vec
x_{\parallel})$. Mathematically, in general, the relation
(\ref{threlation}) gives $T$ in terms of hypergeometric functions of
$H$.

The field redefinition yields
\begin{equation}\label{sstatrad} S_{\rm probe}^{\rm
(static,radial)} = -\tau \int d^{p+1}x_\parallel {V(T)}\left[
\sqrt{1+{\bar{g}}^{\alpha\beta}\partial_\alpha T \partial_\beta T}
\pm\zeta \right]
\end{equation}
where the indices run only over $\alpha,\beta=0,\ldots,p$ and the
only nontrivial component of ${\bar{g}}$ is the time component. In
this probe action (\ref{sstatrad}) for radial-only motion, the
tachyon potential is given implicitly by
\begin{equation}\label{vofthoft}
V(T)\equiv {\frac{1}{H(T)}} \,.
\end{equation}
To obtain the explicit form of $V(T)$, it is necessary to invert the
$T(H)$ expression (\ref{threlation}) to get $H(T)$. For D$p$-brane
geometries where the probe is close to the stack of source branes
(e.g. in the decoupling limit in the string theory context)
\begin{equation}\label{largeH}
H-1 \gg 1\,,
\end{equation}
the above inversion can be done relatively easily. We assume this
holds in the following.

For the general case for source branes (BPS or not), with the
definition
\begin{equation}\label{alphadef}
\alpha\equiv {\frac{2(7-p)}{(5-p)}}
\end{equation}
we obtain the exact expression for $T(r)$ in terms of hypergeometric
functions:
\begin{equation}\label{thyper}
T(r)={\frac{(5-p)}{2R}} H(r)^{1/\alpha}\
{}_2F_1\left({\frac{1}{2}},{\frac{1}{\alpha}};
1+{\frac{1}{\alpha}};1-K(r)\right)
\end{equation}
Mathematically, this expression is monotonic in $r$, and is
convergent for $K(r)\in[0,1]$ (including at the endpoints) because
$\alpha>0$.

When the probe is far from the horizon, in the sense that $K(r)\sim
1^-$, the series expansion of the hypergeometric function can be
employed; this is obviously also valid when there is no horizon (BPS
source branes).  We then get
\begin{equation}\label{powerlawT}
\left. T\right|_{\rm bps} ={\frac{(5-p)}{2R}}
\left({\frac{R}{r}}\right)^{2/(5-p)}
\end{equation}
Hence, large-$T$ corresponds to small coordinate distance from the
source branes.  Since we are considering the large-$H$ limit
(\ref{largeH}), this is always our regime of interest. The potential
for the tachyon
\begin{equation}\label{powerlaw}
\left. V(T)\right|_{\rm bps} =
\left({\frac{2R}{(5-p)}}T\right)^{-\alpha}
\end{equation}
exhibits {\em power-law runaway behavior}.

In the non-BPS case, in order to obtain the near-horizon $K(r)\sim
0^+$ expansion of (\ref{thyper}), we use a standard identity for
hypergeometric functions to get
\begin{eqnarray}\label{thnonbps}
{}_2F_1 \left({\frac{1}{2}},{\frac{1}{\alpha}};
1+{\frac{1}{\alpha}};1-K\right)&=&
{\frac{\Gamma(1+1/\alpha)\sqrt{\pi}}{\Gamma(1/2+1/\alpha)}} \,
{}_2F_1(1/2,1/\alpha;1/2;K) \nonumber\\ &&-
2{\frac{\Gamma(1+1/\alpha)}{\Gamma(1/\alpha)}}\sqrt{K}\,
{}_2F_1(1/2+1/\alpha,1;3/2;K)
\end{eqnarray}
and then, again, employ the series expansion. The upshot is that, in
the near-horizon limit,  the entire expression is finite all the way
to the horizon. The second term in (\ref{thnonbps}) is subleading at
small-$K$, while the first term has a coefficient as a ratio of two
finite $\Gamma$-functions. Amusingly, then, the non-BPS expressions
for $T(H)$, and hence $V(T)$, end up being identical to the BPS case,
up to a finite numerical multiplicative constant. This system
therefore exhibits the same qualitative physics, as far as the
runaway tachyon potential is concerned. The physically relevant
qualitative difference for non-BPS branes is, of course, that the
auxiliary metric on the brane worldvolume,
$({\bar{g}}_{\alpha\beta})={\rm{diag}}(-K(T),1,\cdots,1)$ features a
non-trivial time slowdown near the horizon. In the following, we will
choose to stick to the BPS case, because it can be solved exactly in
an analytic asymptotic series expansion.

\section{Free-streaming approximation}\label{sec:freestream}
The equation of motion for the tachyon field follows from the action
(\ref{sstatrad})
\begin{equation}\label{motion}
\Box T - {\frac{\partial_\mu \partial_\nu T}{1 +
\partial_\alpha T \partial^\alpha T}} \, \partial^\mu T \partial^\nu T
- {\frac{V_{,T}}{V}} \left( 1 + \zeta\sqrt{1+\partial^\alpha T
\partial_\alpha T} \right) = 0 \ .
\end{equation}
This equation is an example of a non-linear, partial differential
equation which, as we will show, admits a relatively simple,
general, inhomogeneous solution.

The energy density of the tachyon field $\rho=T_{0}^0$ is
\begin{equation}\label{rho}
\rho =\frac{V(T)}{\sqrt{1+\partial_{\mu}T \partial^{\mu}T}} \,
\dot{T}^2 + V(T) \, \sqrt{1+\partial_{\mu}T \partial^{\mu}T} ,
\end{equation}
where $V=V(T)$ is given above in (\ref{powerlaw}). As has been
observed in \cite{fks} by solving (\ref{motion}) numerically, if we
define an operator $P(T)=1+\partial_\mu T\partial^\mu T$ the field
$T$ rapidly approaches to a regime in which $P(T) \approx 0$.

Importantly, this means that in the same regime, {\it the
Ramond-Ramond coupling is essentially irrelevant} physically. In
particular, neither the $\pm$ sign (whether the probe is a brane or
an anti-brane) nor the size of $\zeta$ (the non-extremality
parameter) matters! This conclusion is somewhat surprising, naively,
because one would think that the Ramond-Ramond fields would be very
important near the source branes. The point is that the gauge fields
become essentially irrelevant for this test-brane falling close in to
the source branes, because of a strong kinetic suppression. We will
nonetheless keep track of the R-R contribution to the theory on the
probe.

A summary of the free-streaming tool of \cite{fks} goes as follows.
In the leading approximation,
\begin{equation}\label{zero}
T(X^{\mu}) \approx S(X^{\mu}) \ ,
\end{equation}
where $S$ satisfies the equation
\begin{equation}\label{HJ}
\dot S^2-\left({\vec \nabla}_{x} S\right)^2=1 ,
\end{equation}
where we use $t=X^0$ , dot is $\partial_t$ and the spatial
derivatives are with respect to the $p$ spatial coordinates $\vec x
=X^{\mu}$ on the brane. This equation is the Hamilton-Jacobi equation
for the evolution of the wave front function of free steaming massive
relativistic particles. In this particle description, at some initial
time $t_0$ we can label the position of each particle with a vector
$\vec{q}$; equivalently, we can say that $\vec{q}$ parameterizes the
different particles. The initial (covariant) velocity of the particle
is given by $\partial^{\mu} S_0$. If we further define the proper
time $\tau$ along each particle's trajectory, we can switch from
coordinates $(t, \vec{x})$ to $\left(\tau(t, \vec x), \vec q(t, \vec
x)\right)$ and obtain an exact parametric solution to (\ref{HJ})
\cite{fks}
\begin{eqnarray}\label{character}
\label{ssolutionx}
\vec{x} &=& \vec{q} - \vec{\nabla}_{\vec q}S_0 \tau \nonumber\\
\label{ssolutiont}
t &=& \sqrt{1 + \vert \nabla_{\vec q}S_0 \vert^2} \, \tau \nonumber\\
\label{ssolutions}
S &=& S_0 + \tau.
\end{eqnarray}
Notice that the coordinates $(\tau,{\vec{q}})$ are the proper
coordinates of the probe brane.  The interpretation of the solution
(\ref{character}) is very simple and intuitive. It tells us that the
field $S$ propagates along the trajectories of the massive
relativistic particles, growing linearly with proper time. The slope
of each characteristic depends only on the initial gradients of $S_0$
on that characteristic.

In the free streaming approximation, the denominator of the first
term in the tachyon energy density expression (\ref{rho}) is zero,
because the leading term in $\dot{T}^2$ is unity. Given this fact, it
is clear that in order to compute the energy density near the
asymptotic runaway regime, it is necessary to go beyond the free
streaming approximation (\ref{HJ}). In the next section, we compute
the energy density of the tachyon by finding corrections beyond the
free streaming approximation by using a working ansatz in an
asymptotic series expansion.

\section{The full solution}\label{sec:fullsoln}
We are interested in the energy density: something the
free-streaming approximation cannot give us. So let us begin here
by describing the energy density qualitatively. Looking at
equation (\ref{rho}) we see that the potential pieces are growing
small, as an inverse power-law, as are the arguments of the square
roots. The second term in (\ref{rho}) will thus rapidly become
irrelevant. What we need consider is only the competition in the
first term in (\ref{rho}) between the small potential in the
numerator and the kinetic square-root term in the denominator, all
evaluated near the runaway asymptotic region of the potential. In
a previous work \cite{felderkofman}, the asymptotic series
solution for the exponentially decaying potentials was proposed to
be $T(x^{\mu})\approx S+\sum_{n=0}^{\infty}f_{n}e^{-(n+1)S}$. Here
we generalize this result by conjecturing the following asymptotic
expansion for the field  T\footnote{One can try asymptotic series
with respect to other of $S$. However the final answer shall be
independent of the choice.}
\begin{equation}
T(x^{\mu})\approx S+\sum_{n=0}^{\infty} f_{n}S^{-(n+1)\gamma}
\end{equation}
In the next to the leading approximation, we have
\begin{equation}\label{first}
T(X^{\mu}) \approx S+ f_1G_1 \ ,
\end{equation}
where $G_1$ is of the following form:
\begin{equation}
G_1=S^{-\gamma},
\end{equation}
where $\gamma >0$ and
\begin{equation}\label{choice}
\alpha=\frac{\gamma +1}{2}
\end{equation}
Plugging the expansion (\ref{first}) into equation (\ref{motion})
and keeping only terms linear in $f_{1}$, gives the following
equation for the function $f_1$
\begin{eqnarray}\label{f1motion}
-S\partial^\mu S \partial^\nu S (\partial_\mu
\partial_\nu f_1) + \left(2S\Box S - 2\gamma+2\alpha\right)
\partial^\mu S(\partial_\mu f_1)+ &&
\nonumber\\
\big[(2\alpha\gamma-\gamma(\gamma+1))S^{-1}+2\gamma\Box S\big]f_1 &&=
0 \ ,
\end{eqnarray}
where $\Box S=\partial_\nu \partial^\nu S=-\ddot S + {\nabla_{\vec
x}}^2 S$. This equation can be dramatically simplified by changing
from ($t$, $\vec{x}$) coordinates to ($\tau$, $\vec{q}$)
coordinates, as defined by the characteristics of $S$ in
(\ref{ssolutions}). In these coordinates, $f_1$ has no spatial
derivatives and equation (\ref{f1motion}) reduces to
\begin{equation}\label{f1}
Sf_{1,\tau\tau}+ \left(2S\Box S-\gamma+1\right) f_{1,\tau}-
2\gamma\Box S \, f_1 = 0,
\end{equation}
where using (\ref{choice}) we have got rid of the term
proportional to $S^{-1}$. Note that $\Box S$ can be calculated
either with respect to ($t$, $\vec{x}$) or ($\tau$, $\vec{q}$)
coordinates. We can further simplify this equation by introducing
a new variable $y$
\begin{equation}\label{y}
y=\gamma f_1 - Sf_{1,\tau} \
\end{equation}
The second order differential equation (\ref{f1}) can be rewritten
in terms of $y$ satisfying a {\it first} order differential
equation
\begin{equation}
y_{,\tau} + 2 \Box S \, y = 0,
\end{equation}
which can be immediately integrated to give $y(\tau, \vec q) =
y_0(\vec q) \exp\left(-2 \int^{\tau} d\tau' \, \Box S \right) $.
From this and (\ref{y})
\begin{equation}\label{fone}
f_1(\tau, \vec q)=f_{1i}(\vec q) \, S^{\gamma} \,
\int^{\tau}d\tau'S(\tau',\vec q)^{-\gamma-1}e^{-2 \int^{\tau'}
d\tau'' \, \Box S} \ .
\end{equation}
(The integration constant that one gains after integrating
(\ref{y}) is irrelevant as it can be absorbed into $S$.)

To proceed further we need to estimate (\ref{fone}). It can be
shown that $\exp(-2\int^{\tau}d{\tau'}\Box S)$ is a polynomial of
order $2p$ with respect to $\tau$ such that $f_1G_1$ is small in
the asymptotic region. We start by referring back to the
free-streaming parametric solution (\ref{character}). Using the
fact that $T\simeq S + f_1 G_1$, this shows the validity of the
asymptotic series expansion (\ref{first}) for runaway power-law
potentials.We discuss second-order corrections in the Appendix A;
these are also shown to preserve the validity of the asymptotic
series expansion. Comments on arbitrary runaway potentials can be
found in the Appendix B.

To compute the energy density, we need more information. Also, proper
coordinates $(\tau,{\vec{q}})$ are not so easily interpreted in the
spacetime picture, where the brane position in static gauge is given
by $X^\alpha$.  Therefore, let us compute the Jacobian that
transforms $(\tau, \vec{q}) \to (t, \vec{x})$.  To see its role, let
us use the following trick. Plugging the expansion (\ref{first}) into
the energy density (\ref{rho}) we find $\rho \approx
1/\sqrt{2\left(\gamma f_1+S\partial_\mu S
\partial^\mu f_1 \right)}$. The denominator here is none other than
$\sqrt{2y}$.  Therefore,
\begin{equation}\label{energy}
\rho \approx {\frac{1}{\sqrt{2 y}}} = \rho_0(\vec q) \,
\exp\left({\int^{\tau} d\tau' \Box S } \right)\ .
\end{equation}
This is precisely the expression for the energy density of free
streaming relativistic particles which obey the relativistic
continuity equation $\partial_\mu S \left( \rho \, \partial^\mu S
\right)=0$. In fact, (\ref{energy}) is the solution of this
continuity equation in the coordinates $(\tau, \vec{q})$. However,
there is another form of the energy density, which is equivalent to
(\ref{energy})
\begin{equation}\label{energy1}
\rho =\rho_0(\vec q) /\left| \frac{\partial \vec x}{
\partial \vec q } \right|\ ,
\end{equation}
where the denominator is the Jacobian of the $\vec x \to \vec q$
transformation.  Note that we can compare expressions (\ref{energy})
and (\ref{energy1}) for $\rho$ to find a precise expression for $\Box
S$ and then, via (\ref{fone}), for $f_1$. Indeed, from conservation
of the energy density in a differential volume we have $d^p \vec q
\rho_0(\vec q)=d^p \vec x \rho$, which leads to the formula
(\ref{energy1}).  It is now straightforward to calculate the Jacobian
from the formulas (\ref{character}). In $p$ dimensions it is a
polynomial in $\tau$ of order $p$.  This can be seen by using
${\vec{x}}={\vec{q}}-\tau\nabla_q S_0$. We can define the symmetric
tensor $D^a_b\equiv \nabla^{q_a}\nabla_{q_b} S_0$; then the Jacobian
can be written in terms of the eigenvalues of $D$,
$\left\{\lambda_i,i=1\ldots p\right\}$, as the product
$(1-\lambda_1\tau)\ldots (1-\lambda_p)$. Using (\ref{energy1}) for
$\rho$, it therefore becomes clear that there is a critical proper
time where the energy density blows up. In detail, at leading order
in the asymptotic series the energy density is given by
\begin{equation}\label{energyblowup}
\rho = \rho_0(\vec{q}) \prod_{i=1}^p {\frac{1}{(1-\lambda_i\tau)}}
\end{equation}
Corrections to this expression are suppressed by powers of $S$.

A similar behavior in the context of exponential potentials was
observed in \cite{felderkofman}.  We now discuss this phenomenon
in our new context of power-law runaway potentials, relevant in
string theory for D-brane motion, in the next section.

\section{Interpretation and Discussion}\label{sec:discussion}
Consider the probe brane or anti-brane in the background geometry.
Suppose the generic embedding is not exactly located in the
hyperplane $\vec x_{\parallel}$, but rather given by the distance
function $r=r(x_{\parallel})$. Then the following picture emerges of
how the brane is moving in the background geometry. First, very
quickly the brane hypersurface becomes moving in the regime of free
streaming, when each point of the brane is ballistically propagating.
By very quickly we mean that the test brane is in the vicinity of the
near-horizon throat. Simple formulas (\ref{character}), resembling
the Huygens principle of geometric optics, are describing the
evolution of the brane hypersurface $T$; in other words, its radial
position $r$. Geometrical rules and pictorial illustrations of the
free streaming approximation are given in detail in the previous
papers \cite{fks,felderkofman}.

\begin{figure}
\begin{center}
\epsfxsize=.75\columnwidth \epsfbox{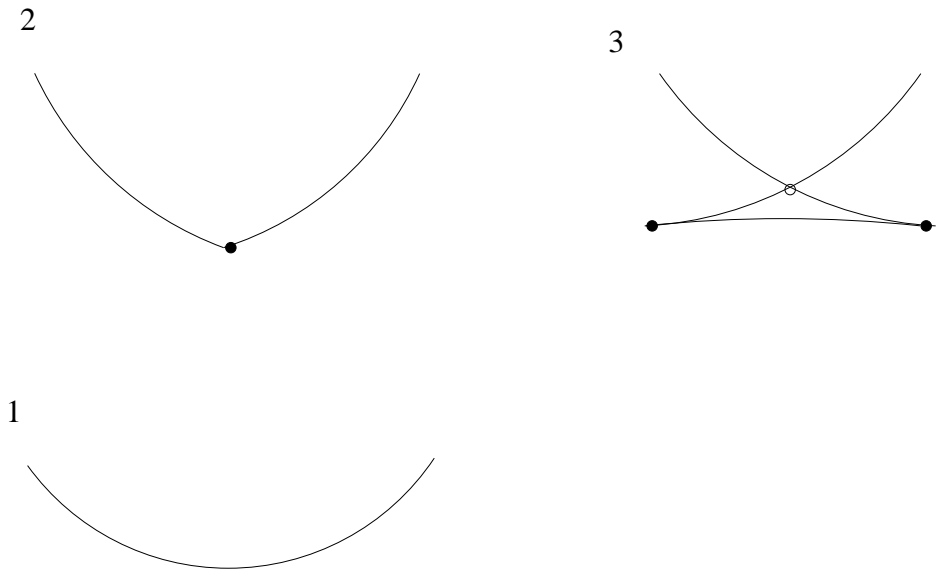}
\caption{Folding of the brane segment from initial profile 1 through
the instance 2 of the first caustic formation to typical multiple
valued configuration 3.} \label{fig:caus}
\end{center}
\end{figure}

The most important feature of brane evolution we see here is
unbounded bending of its hypersurface, in other words the growth of
its inhomogeneities. A simple rule stands, that the brane regions
around local maxima of $T$ are getting flatter while regions around
minima are sharpening. This illustrated by the Figure~\ref{fig:caus}
which shows edge-on profile of the folding brane segment around the
minima of $T$ with the most dramatic reconstruction. At some moment
the brane profile acquires a discontinuity where its second
derivative blows up. This corresponds to caustic formation. Formally,
if we allow the free streaming approximation to work further, the
brane hypersurface is self-crossing, and $T$ goes multi valued as
shown on the figure.

One may think that the multi-valuedness of the field $T$ is related
to the failure of the parametrization of the world-volume of the
probe D$p$-brane in static gauge. We argue that brane self-crossing
is real, and accompanied by local blowup of the stress-energy tensor.
Indeed, we went beyond the free-streaming approximation to describe
evolution of $T$. While next-to-leading order terms in the series
$T\simeq [S+ {\cal{O}}(1/S^n)]$ have essentially negligible
correction to the free-streaming propagation of $T$, they are fully
responsible for the stress-energy tensor. However, the very simple
and transparent expression for stress-energy tensor
(\ref{energy},\ref{energy1}) shows that at a certain critical time
and location the energy density becomes singular. The formula
(\ref{energy},\ref{energy1}) is given in terms of proper time and
proper brane coordinates, and does not relate to the static gauge. We
conclude that formally the effective DBI+WZ theory for generic
inhomogeneous solutions for radial probe-brane motion leads to
multivalued solutions, with caustics where the stress-energy tensor
turns singular.

Certainly, the effective field theory description of the brane
dynamics is limited by finite field gradients and stress energy
components. Therefore, we have to discount the formal results beyond
the points where the field gradients are large (say, larger than
unity in string units, which we used here). {\it What happens with
the probe brane around a region about to develop a caustic?} Here, we
only partly address the issue and suggest some directions for future
thoughts. One possibility is related to the classical physics of
extended objects with sharp kink-like features. Sharp restructuring
of the brane hypersurface may lead to significant radiation similar
to gravitational wave flashes from cosmic string kinks. This would
result in local dissipation of the stress-energy into classical
radiation. Another possibility is better couched in terms of string
theoretic physics. Suppose the multi-valued configuration as shown in
Figure~\ref{fig:caus} begins to form. The horizontal segment of brane
in configuration 3 has orientation opposite to that of the rest of
the brane.  That is to say, this segment now effectively describes a
piece of anti-brane. Brane-antibrane segments in close proximity are
unstable due to the open-string tachyon - corresponding to
fundamental string stretched between the segments - which results in
their annihilation.

Next, we discuss implication of our effect for cosmological
applications.

The inter-brane moduli field $r$ is often exploited as an inflaton in
the string theory constructions of the brane inflation. We mention
one of the realization of this idea which is close to the setup of
paper, namely the inflationary model of \cite{evaat}. The high
dimensional construction includes a D3 brane freely falling in the
background of five-dimensional anti de Sitter geometry generated by a
large stack of source D3-branes. The radius $r$ plays the role of the
moduli field in the effective four-dimensional description. Under
certain conditions, again, in the 4d effective description of'
gravity plus that moduli field inflationary regime can be realized.

One of the most important prediction of inflation is generation of
scalar metric perturbations, usually associated with quantum
fluctuations of the inflaton. Therefore we are interested in spatial
inhomogeneities of the moduli field. In the context of our setup,
these correspond to the spatial inhomogeneities of the brane
embedding.

As we learned, even at the classical level brane inhomogeneities are
unstable and grow. It is therefore interesting to consider combining
probe-brane dynamics with inflation in $3+1$ dimensions. At the level
when 4-dimensional gravity is introduced phenomenologically
(corresponding to adding an Einstein-Hilbert term in the action by
hand), in the DBI effective brane action we just have to introduce
four-dimensional (quasi-) de Sitter geometry. In the equation of
motion for $T$, this will just lead to the friction term $3H \dot T$
and redshifting of all spatial gradients $\partial_{\mu} \to {1 \over
a} \partial_{\mu} $. We can conjecture that apparently folding of the
brane will be suppressed at scales of the de Sitter radius $1/H$.
However, it can persist at larger scales.  A more quantitative
description of this is left for the future.

\section*{Acknowledgements}

The authors wish to thank Gary Felder and Eva Silverstein for useful
discussions. LK and AWP also wish to acknowledge useful
conversations with various participants in the BIRS ``New Horizons
in String Cosmology" meeting in June 2004.

Financial support from an Ontario Graduate Scholarship (OS), NSERC
of Canada (LK,AWP,OS), CIAR (LK,AWP), and the Alfred P. Sloan
Foundation (AWP) is acknowledged.

\section*{Appendix A}
We can use a trick to derive the second-order terms in the
perturbation about free streaming. Namely, we write
\begin{equation}\label{kgtrick}
f_1 G_1 \rightarrow G_1 (f_1 + f_2 G_2)
\end{equation}
where $G_2$ is a  power-law function $G_2\propto S^{-\lambda}$, whose
index $\lambda$ to be determined.

The above trick contains nontrivial information. Indeed, for any
$n$-th order perturbation $f_n$ we will need to use the equation of
motion for the lower-order perturbation(s), which in general yield
an equation for $f_n$ which is not identical to the equation for
$f_1$. Let us see this explicitly. The equations give
\begin{eqnarray}
(\partial_\mu\partial_\nu f_2)(\partial^\mu S) (\partial^\nu S)S -
(\partial_\mu f_2)(\partial^\mu
S)\left[(2\alpha-2\gamma-2\lambda)+2S\Box S
\right] && \nonumber\\
f_2 \left[2(\gamma+\lambda)\Box S -
{\frac{\lambda(\lambda+\gamma)}{S^2}} \right] &&=0
\end{eqnarray}
Changing coordinates to $(\tau,{\vec{q}})$ gives
\begin{equation}
Sf_{2,\tau\tau} + \left[2S\Box S +
(1-\gamma-2\lambda)\right]f_{2,\tau} - \left[2(\gamma+\lambda)\Box S
- {\frac{\lambda(\lambda+\gamma)}{S^2}} \right] f_2 =0
\end{equation}
Defining
\begin{equation}
{\hat{y}} \equiv -S f_{2,\tau} G_2 + (\gamma+\lambda)f_2 G_2
\end{equation}
we find that again the equation becomes integrable:
\begin{eqnarray}
f_2 = && f_{2i}({\vec{q}}) \exp\left(\int^\tau d\tau'
{\frac{(\gamma+\lambda)}{S(\tau')}} \right) \times\nonumber\\
&& \left\{ \int^\tau d\tau'' \exp\left(
-\int^{\tau''}d\tau'{\frac{(\lambda+\gamma)}{S(\tau')}}
-2\int^{\tau''}d\tau' \Box S \right)
{\frac{1}{S(\tau'')^{-\lambda+1}}} \right\}
\end{eqnarray}
At long-time, $S\sim\tau$, and this equation can be integrated
easily. It is then clear that $f_2G_2$ is of the same order as $f_1$.
In other words, the extra piece just renormalizes $f_1$.  This fact
holds true for {\em any} choice of $\lambda$.

We can obtain higher-order equations in perturbation theory by
repeating our trick iteratively: substituting $f_{n-1}\rightarrow
G_{n-1}(f_{n-1}+f_nG_n)$ and using lower-order equations of
motion. We conjecture that at general order additional
perturbations either correspond to a reparametrization of $S$, or
a renormalization of lower-order terms, or pieces which are down
by additional powers of $S$ in an asymptotic expansion.

\newpage
\section*{Appendix B}
It is interesting to ask if the analysis at first order beyond the
free-streaming approximation can be done for {\em arbitrary}
runaway tachyon potential. Indeed, we show how to do this here.

We start by writing
\begin{equation}
T(x^{\mu})\approx S + \epsilon (\tau,\Vec q)
\end{equation}
Next, we plug this expression into the general nonlinear equation
of motion (\ref{motion}) for the tachyon.

In an asymptotic series expansion, $\epsilon$ should be much smaller
than $S$, this will prove to be a self-consistent approximation.
Keeping only lowest-order terms in $\epsilon$, and using the equation
of motion for $S$, we find a surprise: the equation for $\epsilon$
involves only derivative terms.

The next simplification occurs upon changing to proper coordinates
$(\tau, \vec q)$. The resulting equation of motion for $\epsilon$
becomes
\begin{eqnarray} -\epsilon_{,\tau \tau}+\epsilon_{,\tau}\left[-2\Box
S+2\left.\left(\frac{V_{,T}}{V}\right)\right|_{T=S} \right]=0
\end{eqnarray}
which can be easily integrated (twice) to give
\begin{equation}
\epsilon(\tau,\vec q)=\epsilon_{i}(\vec q)\int^{\tau}
d{\tau''}\exp\left(2\int^{\tau''}d{\tau'}
\frac{V_{,S}}{V}\right)\exp\left(-2\int^{\tau''} d{\tau'} \Box
S\right)
\end{equation}

In the long-time limit, or when initial inhomogeneities are small
(which is valid here) we can do the integral involving the potential
to give
\begin{equation}
\epsilon(\tau,\vec q)=\epsilon_{i}(\vec q)\int^{\tau} d{\tau''}
\left[V\left(S(\tau'')\right)\right]^2\exp\left(-2\int^{\tau''}
d{\tau'} \Box S\right)
\end{equation}
As is clear from the above equation, in the long-time limit, the
suppression of the next-to-leading term is controlled by the square
of the tachyon potential, as was suggested in section 4. For runaway
potentials, this is always small in the regime of validity of the
asymptotic series expansion.  In particular, note that our results
here agree with the case of the power-law worked out in the body of
the paper.

The energy-momentum tensor can then be written as
\begin{equation}\label{rh}
\rho(\tau, \vec q)\approx \frac{V(S)}{\sqrt{\Delta}}
\end{equation}
where\footnote{In proper coordinates, of course, $\partial^\beta S
\partial_\beta \epsilon=-2\epsilon_{,\tau}$.}
\begin{equation}
\sqrt{\Delta}\approx
\sqrt{2\partial^{\beta}S\partial_{\beta}\epsilon} + ({\rm higher\
order\ terms}) \,.
\end{equation}
We have $\partial^{\beta}S\partial_{\beta}\epsilon\sim
V^2\exp(-2\int^{\tau}d{\tau'}\Box S)$. Substituting this into
equation (\ref{rh}) we obtain equations (\ref{energy}),
(\ref{energy1}). Thus we find out that equations (\ref{energy})
and (\ref{energy1}) for the energy density are valid even for
arbitrary runaway potentials. This quantity $\sqrt{\Delta}$
controls the degree of kinetic suppression, of e.g. the
Ramond-Ramond terms in the D-brane probe action.



\end{document}